\documentclass[a4paper]{jpconf}
\usepackage{graphicx}
\begin{document}
\title{Electroweak Symmetry Breaking in the E$_6$SSM}

\author{\underline{Peter Athron}$^{1,3}$, SF King$^2$, DJ Miller$^1$, S Moretti$^2$ and R Nevzorov$^{1}$
}
 \address{$^1$Department of Physics and Astronomy, University of Glasgow, Glasgow G12 8QQ, UK} 

 \address{$^2$School of Physics and Astronomy, University of Southampton, Southampton, SO17 1BJ}

\ead{$^3$p.athron@physics.gla.ac.uk}
\begin{abstract}
The Exceptional Supersymmetric Standard Model (E$_6$SSM) is an E$_6$ inspired model with an extra gauged U(1) symmetry, which solves the $\mu$-problem in a similar way to the NMSSM but without the accompanying problems of singlet tadpoles or domain walls. It predicts new exotic particles at the TeV scale.  We investigate the Renormalisation Group Evolution of the model and test electroweak symmetry breaking for a selection of interesting scenarios with non-universal Higgs masses at the GUT scale.  We present a detailed particle spectrum that could be observed at the LHC. 
\end{abstract}

\section{Introduction}

The E$_6$SSM\cite{King:2005jy,King:2005my} is an E$_6$ inspired model with an extra gauged U(1) symmetry. It provides a low energy alternative to the Minimal (MSSM) and Next to Minimal (NMSSM) Supersymmetric Models. The gauge group is \mbox{SU(3)$\times$ SU(2)$\times$U(1)$_Y\times$U(1)$_N$}, where U(1)$_N$ is defined by U(1)$_N = 1/4$U(1)$_\chi + \sqrt{15}/4$U(1)$_\psi$, with U(1)$_\chi$ and U(1)$_\psi$ in turn, defined by the breaking, E$_6\rightarrow$SO(10)$\times$U(1)$_\psi$ and SO(10)$\rightarrow$SU(5)$\times$U(1)$_\chi$. 

The matter content is based on three generations of complete $27$plet representations of E$_6$ in which anomalies are automatically cancelled. Each $27$plet, $(27)_i$, is filled with one generation of ordinary matter; singlet fields, $S_i$; up and down type Higgs like field, $H_{2,i}$ and $H_{1,i}$ and exotic coloured matter, $D_i$, $\bar{D}_i$. In addition the model contains two extra SU(2) doublets, $H'$ and $\bar{H}'$, which are required for gauge coupling unification.  

The superpotential is $W_{E_6SSM} \approx \lambda_iSH_{1,i}H_{2,i}+\kappa_i S D_i\bar{D}_i+h_t H_u Qt^c + h_b H_d Q b^c + h_{\tau} H_d L \tau^c$, where $H_u = H_{2,3}$ and $H_d = H_{1,3}$ and $S= S_3$ develop vevs, $\langle H^0_u \rangle = v_u$,  $\langle H^0_d \rangle = v_d$ and  $\langle S \rangle = s$. $v_u$ and $v_d$ give mass to ordinary matter via the Higgs mechanism, while $s$ both gives mass to the exotic coloured fields, $\kappa_i S \rightarrow \kappa_i s = m_{D_i}$ and generates an effective $\mu$-term, $\mu_{eff} = \lambda_3 s$. 

Another striking advantage of this model is that the upper bound on the mass of the lightest Higgs is $155$ GeV, significantly larger than in either the MSSM or the NMSSM.     


\section{EWSB in the E$_6$SSM}
Our current work is to test if, following the imposition of some simple high scale assumptions, electroweak symmetry breaking (EWSB) may be radiatively driven in the E$_6$SSM as happens in the MSSM and NMSSM. This is an important test for the model and allows us to predict mass spectra which could be seen at the LHC from a small number of GUT scale parameters.  

Solutions with universal scalar masses at the GUT scale, in addition to universal gaugino masses and universal trilinears, are difficult to find. Relaxing the high scale assumptions to allow non-universal Higgs masses (NUHM) avoids this difficulty. This gives us the NUHM E$_6$SSM GUT scale constraints: universal gauginos ($M_{1/2}$), universal trilinears ($A$), GUT scale Higgs masses ($m_S^{GUT}$, $m_{H_u}^{GUT}$ and $m_{H_d}^{GUT}$) and a common scalar mass ($m_0$) for all other E$_6$SSM scalars. The NUHM E$_6$SSM is an interesting scenario in it's own right, but can also be used as a starting point in the search for solutions with stronger universality assumptions.

Here we present results of the first study into EWSB in this NUHM E$_6$SSM model. As a test case we chose the new E$_6$SSM Yukawas to be $\lambda_3 = 0.6,\, \lambda_{1,2} = 0.46$ and $\kappa_{1,2,3} = 0.162$ at the GUT scale. We also chose $\tan \beta = v_u/v_d = 10$ and $s= 3$ TeV and fixed $v^2 = v_u^2 + v_d^2 = (174~\rm{GeV})^2$.

Next we obtained Renormalisation Group Equation\footnote{Two loop beta functions were used for the susy preserving E$_6$SSM parameters, but the soft susy-breaking parameters were only evolved with one loop RGEs.} (RGE) solutions for soft masses at the electroweak (EW) scale, allowing us to write all low energy soft masses of the E$_6$SSM in terms of the NUHM E$_6$SSM GUT scale parameters, \begin{eqnarray}m_i^2 = \alpha_i M_{1/2}^2 + \beta_i A^2 + \gamma_i A M_{1/2} + \delta_i m_0^2 + \epsilon_i m_S^{GUT\,2} + \rho_i m_{H_u}^{GUT\,2} + \zeta_i   m_{H_d}^{GUT\,2}, \\ A_j = a_j M_{1,2} + b_j A \;\;\;\;\;\;\;\; M_k = c_k M_{1/2},\end{eqnarray}
 where $m_i$ are the scalar masses, $A_j$ the trilinears and $M_k$ the gauginos masses and each mass has it's own set of dimensionless coefficients relating it to the GUT scale parameters.
EWSB breaking constraints fix the three Higgs masses ($m_{H_u}$, $m_{H_d}$ and $m_S$) at the EW scale in terms of the Yukawas, gauge couplings, vevs and $A_{\lambda_3}$. The RGE solution for $A_{\lambda_3}$ can be substituted into this. Since all vevs and susy-preserving parameters were either set to their observed values or already chosen, the Higgs masses can then be written as functions of $A$ and $M_{1/2}$. For example $m_S^2 = p M_{1/2} + q A + l + 1\rm{loop}$, where $p$,$q$ and $l$ are dimensionful coefficients fixed by the vevs and susy-preserving parameters.

\begin{figure}[h]
\begin{center}
\includegraphics[width=148mm,clip=true]{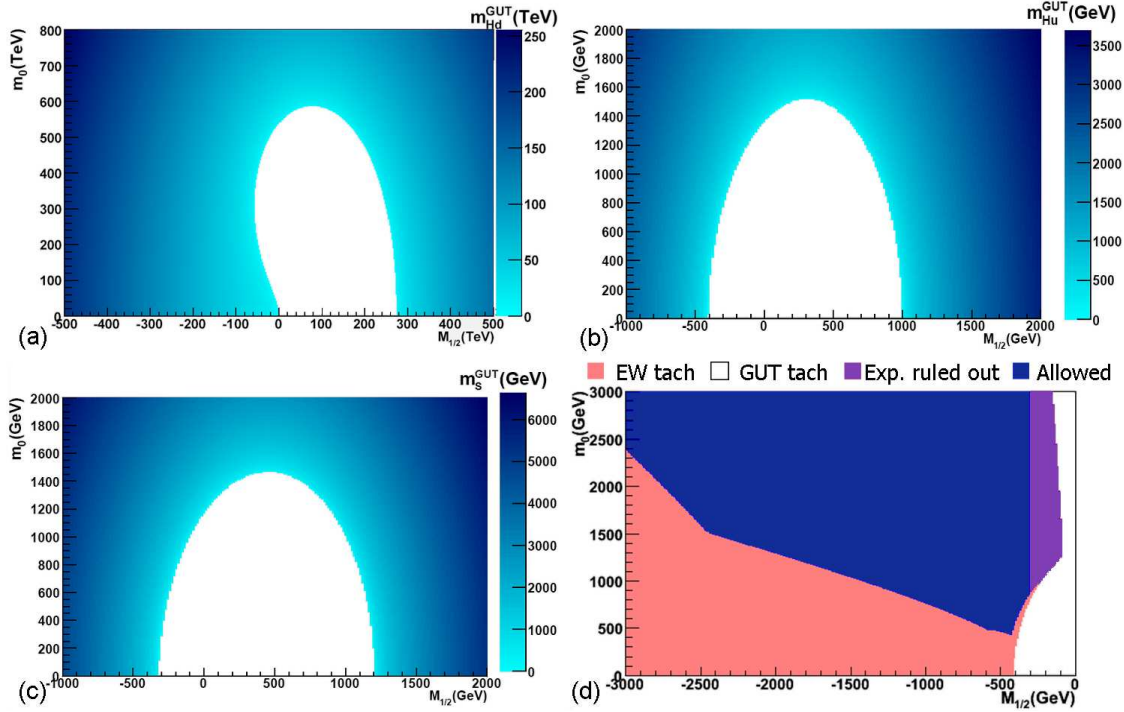}
\end{center}
\caption{  EWSB exclusion plots in the $m_0$, $M_{1/2}$ plane, with $A = -300~\rm{GeV}$. {\it (a)}: $m_{H_d}^{GUT}$ values consistent with the EWSB and NUHM E$_6$SSM boundary conditions. The white region is ruled out since $(m_{H_d}^{GUT})^2 <0$. {\it (b)}: As (a) but for $m_{H_u}^{GUT}$.   {\it (c)}: As (a) but for $m_{S}^{GUT}$\hspace{-1mm}.   {\it (d)}: Full exclusion plot. The blue (black) region shows the allowed region of the parameter space, the purple (dark grey) region is ruled out by searches for the charginos and gluinos, while the pink (light grey) and white regions are ruled out by EW and GUT scale tachyons respectively.     }
\label{EWSB_Exclusion_plots}
\end{figure}

So for each of the Higgs masses we have two constraints which may be equated leaving three low energy constraints and six GUT scale parameters. For all choices of $M_{1/2}$, $A$ and $m_0$ we can then determine what values of $m_{H_d}^2$, $m_{H_u}^2$ and $m_S^2$ are required to satisfy the constraints.

The values obtained for these parameters are shown in Fig.\ref{EWSB_Exclusion_plots}(a-c) with the large white regions ruled out by GUT scale tachyons when $(m_{H_d}^{GUT})^2 < 0$, $(m_{H_u}^{GUT})^2 < 0$ or $(m_{S}^{GUT})^2 < 0$ respectively. From (a) all parameter space accessible at future colliders with $M_{1/2} > 0$ is excluded, but $M_{1/2} < 0$ is valid in our scheme\footnote{Often conventions are chosen such that $M_{1/2} > 0$. We haven't done this here but one can perform the transformation, $A\rightarrow -A, \,M_{1/2}\rightarrow -M_{1/2}, s \rightarrow -s$, to obtain solutions with $M_{1/2}>0$ but with different signs for some Yukawas and vevs.}. (b),(c) exclude some additional points with $M_{1/2} > 0$. Fig.\ref{EWSB_Exclusion_plots}(d) shows the combined exclusion region from (a-c) and also a large region of parameter space ruled out by EW scale tachyons and a small slice ruled out by experiment.

\begin{figure}[h]
\includegraphics[width=90mm,clip=true]{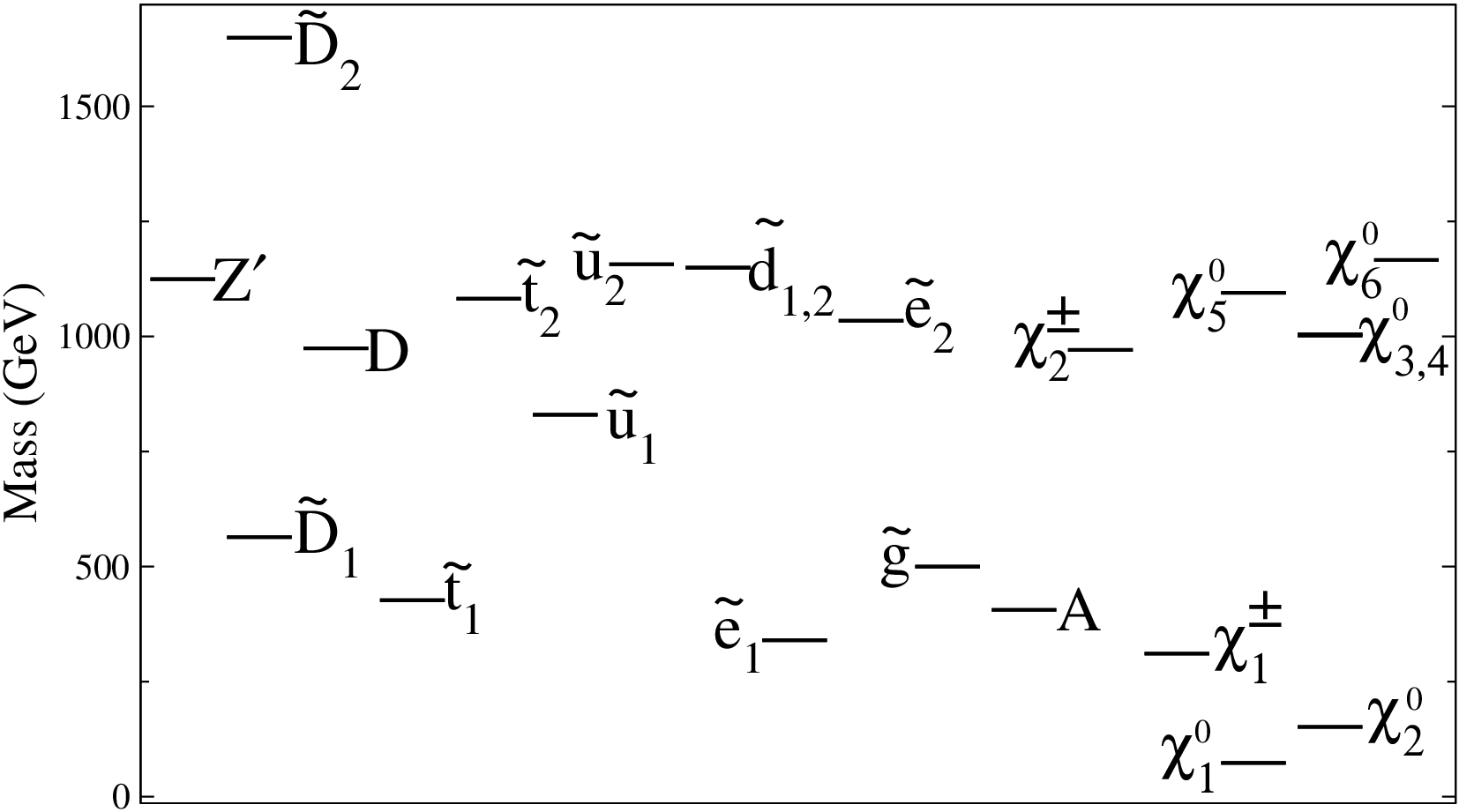}
\begin{minipage}[b]{16pc}\caption{A sample NUHM E$_6$SSM spectrum that could be seen at the LHC. The parameters for this point are \mbox{$m_0 = 600$ GeV}, \mbox{ $M_{1,2} = -500$ GeV}, \mbox{$A = -300$ GeV}, \mbox{ $m_S^{GUT} = 2.39$ TeV}, \mbox{$m_{H_u}^{GUT} = 2.25$ TeV} \& \mbox{ $m_{H_d}^{GUT} = 922$ GeV}} \end{minipage}                    
\label{sample_ESSM_spectrum}
\end{figure}

While EWSB provides strong constraints on the NUHM E$_6$SSM, there is a large volume of parameter space which may be physically realised and could be discovered at the LHC.  A sample spectrum which might be observed is shown in Fig.\ref{sample_ESSM_spectrum}. The presence of the exotic coloured particles ($\tilde{D}_{1,2}$ and $D$) and the new $Z'$ boson at low energies provide interesting experimental signatures for the model.  The gluino is lighter than most of the squarks, which is unusual and phenomenologically interesting as it affects cascade decays. This is a result of the renormalisation group (RG) running of the soft masses in the E$_6$SSM and is typical of points we have looked at in the NUHM E$_6$SSM. The smaller the hierarchy amongst the GUT scale soft parameters, the stronger this effect.  If solutions with fully universal scalar masses at the GUT scale can be found then the gluino will be lighter than all squarks and sleptons.

\section{Conclusions}

For our benchmark choice of E$_6$SSM Yukawas and vevs there were no solutions with all universal scalar masses at the GUT scale. We allowed non universal Higgs masses, and obtained a dramatic improvement. We discovered many spectra that could be seen at the LHC. The RG flow implies that the gluino is often lighter than the squarks. To improve this study we want to include two loop RGEs for gaugino masses, which preliminary results suggest are much more significant than may naively be anticipated, and look for solutions with stronger universality assumptions.

\end{document}